# Optical absorption induced by UV laser radiation in Ge-doped amorphous silica probed by in situ spectroscopy

F. Messina[1], M. Cannas[1], R. Boscaino[1], S. Grandi[2], P. Mustarelli[2]

[1] Physical and Astronomical Sciences Dept. of University of Palermo, Via Archirafi 36, Palermo, Italy
[2] Chemical Physics Dept. of University of Pavia, Viale Taramelli 16, Pavia, Italy

We studied the optical absorption induced by 4.7eV pulsed laser radiation on Ge-doped a-SiO$_2$ synthesized by a sol-gel technique. The absorption spectra in the ultraviolet spectral range were measured during and after the end of irradiation with an *in situ* technique, evidencing the growth of an absorption signal whose profile is characterized by two main peaks near 4.5eV and 5.7eV and whose shape depends on time. Electron spin resonance measurements performed *ex situ* a few hours after the end of exposure permit to complete the information acquired by optical absorption by detection of the paramagnetic Ge(1) (GeO$_4$)$^-$ and Ge-E' ($\equiv$Ge•) centers laser-induced in the samples.

**1 Introduction.** Many studies have evidenced that laser exposure of Ge-doped silica is able to induce transparency loss, photosensitivity and optical nonlinearity in the material, related to generation and conversion of point defects triggered by laser light and highly interesting from an applicative point of view [1-2]. These processes have been extensively studied with multiple spectroscopic techniques, such as optical absorption (OA) and electron spin resonance (ESR) but many questions still remain open [3-9]; in particular, their kinetics during and after laser irradiation has not been sufficiently investigated, due to the lack of appropriate *in situ* techniques, which allow to evidence transient features of the spectroscopic signals inaccessible to *ex situ* observation. Also, the attribution of the absorption bands to the known Ge-related defects is still an open problem.

In this work, we studied the OA induced by 4.7eV pulsed laser radiation on Ge-doped a-SiO$_2$ synthesized by a sol-gel technique. OA spectra in the UV (3-6eV) were measured during and after the end of irradiation with an *in situ* technique, revealing the growth of an absorption profile featuring at least two bands peaked near 4.5eV and 5.7eV and the partial bleaching of the native 5.1eV absorption. A time dependence during and after the end of irradiation of both intensity and shape of the induced OA is evidenced. Finally, ESR measurements performed *ex situ* a few hours after the end of exposure permit to identify the paramagnetic defects laser-induced in the samples.

**2 Materials and Methods.** Samples of Ge-doped SiO$_2$ with 10$^4$ molar ppm Ge content were synthesized by a sol-gel technique, using silicon tetraethoxide (TEOS) and germanium tetraethoxide (TEOG) as the starting materials [10]. The absorption profile of the as-prepared materials (inset of Fig. 1-A) shows the B$_{2\beta}$ band peaked at (5.13±0.02)eV, with (0.43±0.02)eV full width at medium height (FWMH) and 1.2cm$^{-1}$eV area, due to the portion of Ge impurities arranged in the twofold coordinated form (=Ge••) [11].

The cylindrically shaped samples were irradiated perpendicularly to the lateral surface, at room temperature, by the 4.7eV 4$^{th}$ harmonic of the pulsed radiation emitted by a Q-switched Nd:YAG laser (pulsewidth 5ns, repetition rate 2.5Hz). Laser intensity of I=40mJ/cm$^2$ was measured by a pyroelectric detector.

The absorption profile in the UV (3eV-6eV) was measured *in situ* during and for ~1 hour after the end of irradiation by an optical fiber spectrophotometer based on a D$_2$ lamp source and on a 1200lines/mm grating dispersing on a CCD detector. Electron spin resonance (ESR) measurements were carried out by a Bruker EMX spectrometer working in X band (9.8GHz). We used a 0.1mT modulation amplitude and a 1.6mW microwave power, properly chosen to avoid saturation in detecting the signals of Ge-related paramagnetic defects. Concentration of the defects was calculated by comparison of the doubly-integrated signal with a reference sample where the absolute spin density was known by spin-echo measurements [12].

**3 Results and Discussion.** In Fig. 1 is shown the difference absorption spectrum detected in the sample during irradiation with a sequence of 2500 laser pulses (A) and at different delays after the end of exposure (B). Laser induces the growth of two broad absorption features, peaked around 4.5eV and 5.7eV, whose maximum amplitudes are 1.0cm$^{-1}$ and 1.7cm$^{-1}$ respectively. The peak position of the 5.7eV band moves towards the higher energies during exposure. In the post-irradiation range, the shape of the overall absorption profile undergoes a progressive change. In detail, we observe a reduction of the absorption coefficient around 6eV, an increase around 5.0eV and a shift of the 4.5eV component towards the high energies.

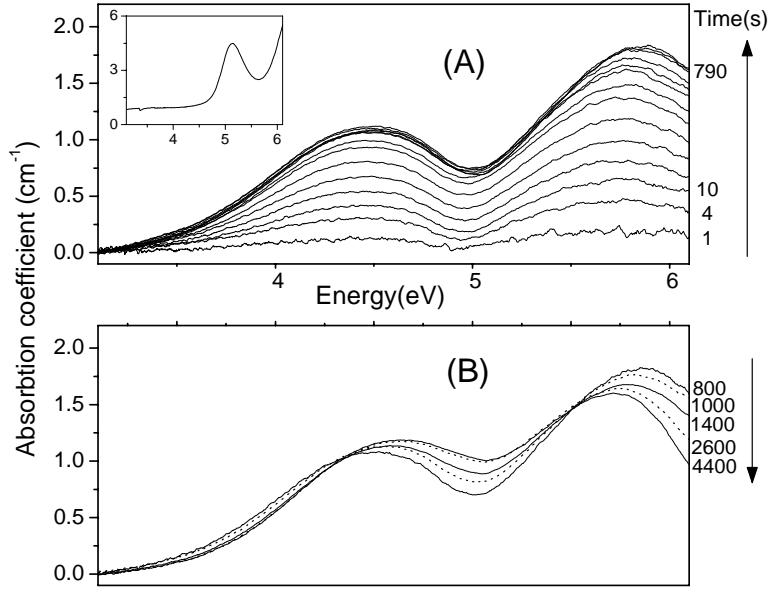

**Fig. 1** Difference absorption spectrum measured *in situ* in a Ge-doped a-SiO$_2$ sample at different times during (A) and after the end of 4.7eV pulsed laser irradiation. Inset: absorption spectrum of the as-grown sample.

To describe quantitatively the kinetics of the OA, we carried out a deconvolution procedure on the spectra. We found that at least 4 gaussian components are required to reproduce the observed profile (inset of Fig. 2-A): the two main peaks around 4.5eV and 5.7eV, a small negative component at 5.0±0.1eV and a band partially falling outside the measured range, whose peak was fixed to 6.1eV. The negative component can be interpreted as the bleaching of the preexisting $B_{2\beta}$ band; hence, in the fitting procedure its FWMH was fixed to the 0.43eV value estimated from the native absorption spectrum. In regards to the widths of the other three components, we found that in they can be fixed to 1.17±0.02eV (4.5eV), 0.85±0.02 eV (5.7eV) and 0.61±0.05eV (6.1eV) in fitting all the spectra. On this basis, the deconvolution procedure yields the 4 areas (Fig. 2-A) and the peaks of the 4.5eV and 5.7eV bands (Fig. 2-B and Fig. 2-C respectively) as a function of time.

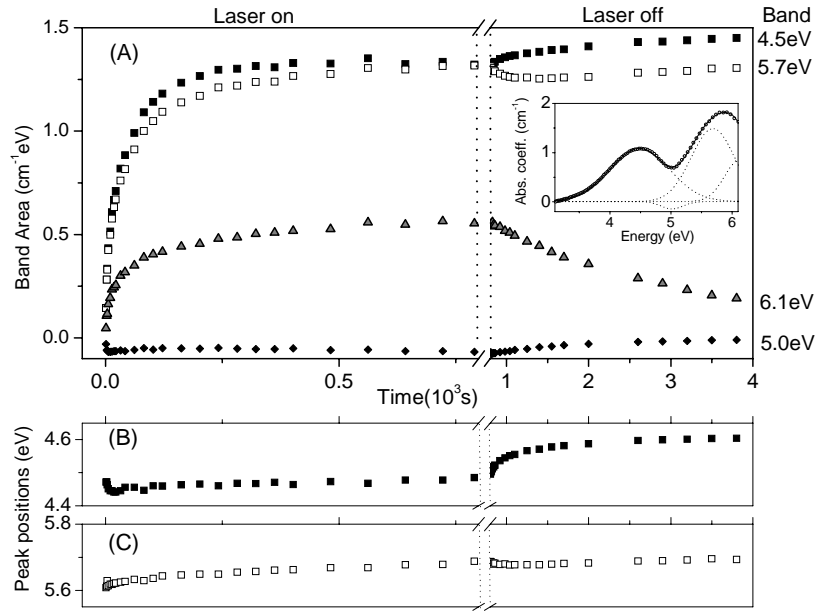

**Fig. 2** Results of absorption profile analysis performed by a 4-peaks deconvolution procedure. Panel (A): areas of the bands as a function of time. Inset of panel (A): example of deconvolution. Panel (B) and (C): peak positions of the 4.5eV and 5.7eV components, respectively.

We see that during irradiation the areas grow to saturation in a few hundreds of seconds. Moreover, the peak position of the 4.5eV component is constant, whereas that of the 5.7eV component moves from 5.61eV to 5.69eV. In the first ~3×10$^3$s of the post-irradiation stage, we observe a kinetics of some of the observed spectral features.

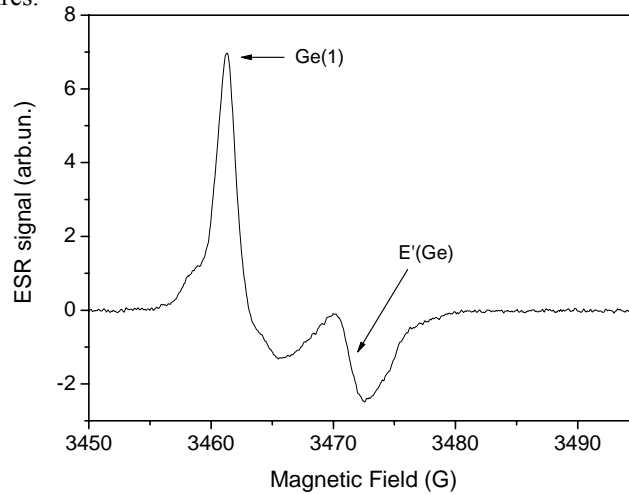

**Fig. 3** ESR spectrum detected in a Ge-doped SiO$_2$ sample a few hours after the end of 4.7eV laser irradiation. The labels indicate spectral features characteristic of two Ge-related paramagnetic point defects.

Indeed, the 6.1eV and the 5.0eV bands are transient and disappear almost completely, this being consistent with the variations of the absorption coefficient around 5eV and 6eV in Fig. 1. Moreover, the peak of the 4.5eV component moves from 4.48eV to 4.60eV.

Finally, we performed ESR measurements on the sample a few hours after the end of exposure. We detect a composite signal, reported in Fig. 3, whose two main components can be attributed to the paramagnetic Ge(1) (GeO$_4$•)$^-$ and E'-Ge (≡Ge•) centers on the basis of comparison with literature and previous work. Neither the typical negative peak of Ge(2) centers (=Ge•) nor the 11.8mT doublet of H(II) centers (=Ge-H•) are detected [3-9,13-14]. The total concentration of paramagnetic defects is $(3.0\pm0.3)\times10^{16}$cm$^{-3}$

Comparison between ESR and OA results leads to discuss the relationship between the several absorption components and the observed Ge-related paramagnetic centers. In detail, present data are consistent with the attribution of the 4.5eV band to the Ge(1) center [4-6,13], whereas we can exclude that the 5.7eV component is due to Ge(2), as put forward by some authors [13], because the ESR signal of the paramagnetic center is not observed; hence, the 5.7eV component must be attributed either to Ge(1) or to some unknown diamagnetic defect. Also, the transient band at E>6eV, which disappears completely in the first ~3×10$^3$s of the post-irradiation stage, seems to be unrelated to Ge(1) and E'-Ge, which at variance are still detected in the material even a few hours after exposure. This contrasts with the common practice of attributing the absorption component at E>6eV observed in irradiated Ge-doped silica to the E'-Ge center [4,6]. The peak drifts of the Ge(1) 4.5eV band and of the 5.7eV bands are symptomatic of a strong inhomogeneity, likely responsible also of the large width of the former band. Similar peak drifts have been observed also upon thermal treatment of X-irradiated Ge-doped silica samples [6]. Finally, the absence of H(II), which is formed by reaction of H$_0$ with twofold coordinated Ge, so being a fingerprint of the presence of diffusing hydrogen [8], permits to exclude that the mobile specie is involved in the observed post-irradiation processes.

**5 Conclusions.** *In situ* OA measurements evidence a complex time-dependence of the absorption shape induced in Ge-doped silica by laser-irradiation, likely conditioned by a strong inhomogeneous contribution to the width of the 4.5eV Ge(1) band. Comparison with ESR data suggests that signals related to unknown diamagnetic centers contribute significantly to the induced OA. More studies may help to clarify the intricate picture emerging from the overall results.

**Acknowledgements**　The authors would like to thank S. Agnello for useful discussions.